\documentclass[12pt]{article}
\usepackage[latin1]{inputenc}
\usepackage{pstricks}
\usepackage{amsmath}
\usepackage{float}
\usepackage{color,soul}
\input epsf
\usepackage{amssymb}
\usepackage{graphicx}
\usepackage{amssymb}
\usepackage{epstopdf}
\usepackage[left=2.3cm,top=3cm,right=2.3cm,bottom=3cm]{geometry}

\newcommand{\be}{\begin{equation}}
\newcommand{\ee}{\end{equation}}
\newcommand{\bea}{\begin{eqnarray}}
\newcommand{\eea}{\end{eqnarray}}
\newcommand{\lb}{\label}

\def \g{\gamma}
\def \ot{\Omega^{\rm tot}_m}
\def \e{\varepsilon}
\def \go {\gamma^{\varepsilon}_{-\infty}}

\def \dg{\gamma_{\Omega}}
\def \ddg{\gamma_{\Omega \Omega}}

\begin{document}
\begin{titlepage}
\title{Global properties of the growth index: \\ 
mathematical aspects and physical relevance}
\author{
R. Calderon$^1$\thanks{email:rodrigo.calderon-bruni@umontpellier.fr}, 
D. Felbacq$^1$\thanks{email:didier.felbacq@umontpellier.fr}, 
R. Gannouji$^2$\thanks{email:radouane.gannouji@ucv.cl}, 
D. Polarski$^1$\thanks{email:david.polarski@umontpellier.fr},
A.~A.~Starobinsky$^{3,4}$ \thanks{email:alstar@landau.ac.ru}
\hfill\\
$^1$ Laboratoire Charles Coulomb, Universit\'e de Montpellier \& CNRS\\
 UMR 5221, F-34095 Montpellier, France\\
$^2$ Instituto de F\'{\i}sica, Pontificia Universidad  Cat\'olica de Valpara\'{\i}so,\\
 Casilla 4950, Valpara\'{\i}so, Chile\\
$^3$ Landau Institute for Theoretical Physics RAS, Moscow, 119334, Russia\\
$^4$ National Research University Higher School of Economics,\\
 Moscow 101000, Russia }
\pagestyle{plain}
\date{\today}

\maketitle

\begin{abstract}
We analyze the global behaviour of the growth index of cosmic inhomogeneities in an isotropic 
homogeneous universe filled by cold non-relativistic matter and dark energy (DE) with an 
arbitrary equation of state. Using a dynamical system approach, we find the critical points 
of the system. That unique trajectory for which the growth index $\gamma$ is finite from the 
asymptotic past to the asymptotic future is identified as the so-called heteroclinic orbit 
connecting the critical points $(\Omega_m=0,~\g_{\infty})$ in the future and 
$(\Omega_m=1,~\g_{-\infty})$ in the past. The first is an attractor while the second is a saddle 
point, confirming our earlier results. 
Further, in the case when a fraction of matter (or DE tracking matter) $\e \Omega^{\rm tot}_m$ 
remains unclustered, we find that the limit of the growth index in the past $\g_{-\infty}^{\e}$ 
does not depend on the equation of state of DE, in sharp contrast with the case $\e=0$ (for 
which $\g_{-\infty}$ is obtained). 
We show indeed that there is a mathematical discontinuity: one cannot obtain $\g_{-\infty}$ by 
taking $\lim_{\e \to 0} \gamma^{\e}_{-\infty}$ (i.e. the limits $\e\to 0$ and 
$\Omega^{\rm tot}_m\to 1$ do not commute).
We recover in our analysis that the value $\g_{-\infty}^{\e}$ corresponds to 
tracking DE in the asymptotic past with constant $\gamma=\g_{-\infty}^{\e}$ found earlier. 
\end{abstract}

\end{titlepage}

\section{Introduction}
The present accelerated expansion rate of the Universe remains an outstanding challenge for 
theoretical cosmology. Despite intensive ongoing activity, the nature of dark energy (DE) 
driving the present accelerated expansion stage (physical, geometrical, or both) and its 
relation to known particles and fields remain unsettled \cite{SS00}. Many DE models inside, 
as well as outside, general relativity (GR) were suggested for this purpose. While the 
increasing accuracy of observations allow to rule out many of them, a large number still 
remains viable. Among the successful DE models, $\Lambda$CDM has a very particular place due 
to its remarkable simplicity: it is based on GR with cold non-relativistic matter as a source, 
and requires only the addition of a (cosmological) constant $\Lambda$ into the Einstein 
equations. However, the attempt to interpret $\Lambda$ in terms of 'vacuum energy' of quantum 
fields requires understanding why its effective energy density is so small compared to all 
other known substances. On the other hand, from the classical point of view, the $\Lambda$CDM 
is intrinsically consistent and its phenomenology serves as a benchmark for the interpretation 
of observational data and comparison to other DE models. Future observations will strongly 
constrain surviving models \cite{WMEHRR13}. It is therefore important to have tools 
characterizing their phenomenology (see e.g. \cite{SSS14}). One such tool is the growth 
index $\gamma$.   

The growth index has a nice property valid for $\Lambda$CDM and more generally for 
non-interacting smooth DE models inside GR \cite{PG07}: up to a small correction depending on 
$\Omega_{m,0}$, its value today $\gamma_0$ is well constrained, $\gamma_0\approx 0.55$. 
In addition, at higher redshifts it is quasi-constant (see also \cite{PSG16}). For example, in 
the presence of a cosmological constant $\Lambda$, $\gamma$ tends to $\frac{6}{11}$ for 
$z\gg 1$ and it departs little from that value even up to the present time.
Its discriminative power is therefore limited for these models. However modified gravity DE 
models can exhibit a strong departure from this behaviour \cite{LC07},\cite{AQ04},\cite{GMP08},
\cite{MSY10}. The growth index offers therefore the possibility to discriminate between DE 
models inside and outside GR, motivating its study in the context of DE models. 
Hence, while the growth index was initially introduced in order to characterize the growth 
of matter perturbations for open Universes \cite{P84}, and later generalized to other models 
inside GR \cite{LLPR91}, interest in the growth index was revived recently \cite{LC07} for the 
assessment of DE models.      

The study of the growth index is also of mathematical interest in its own. A global analysis 
of its dynamics, from deep in the matter era till the future DE dominated stage, often offers a 
better insight on its evolution including low redshift behavior probed by observations 
\cite{CFGPS19}. We will study in details a system with partially unclustered dust-like matter 
(i.e., it could be either ultra-light (axion-like) dark matter, or DE tracking 
dust-like matter, or both) showing interesting connections with results obtained 
earlier for a strictly constant growth index. We will also study the evolution of $\gamma$ 
using the dynamical system analysis. We first review the basic formalism in the next section, 
as well as results and methods from our earlier work \cite{CFGPS19}.     

\section{The growth index}
In this section, we review the basic equations and definitions concerning the growth index 
$\gamma$. 
We consider the evolution of linear (dust-like) matter density perturbations 
$\delta_m =\delta\rho_m/\rho_m$ on cosmic scales. 
Deep inside the Hubble radius they obey the following equation 
\be
{\ddot \delta_m} + 2H {\dot \delta_m} - 4\pi G\rho_m \delta_m = 0~,\label{del}
\ee
where $H(t)\equiv \dot a(t)/a(t)$, resp. $a(t)$, is the Hubble parameter, resp. the 
scale factor of a Friedmann-Lema\^ itre-Robertson-Walker (FLRW) universe filled with 
standard dust-like matter and DE (we neglect radiation at the matter and DE dominated stages). 
We assume that DE remains non-clustered gravitationally at scales at which matter does.

For vanishing spatial curvature, we have for $z\ll z_{eq}$
\be
h^2(z) = \Omega_{m,0} ~(1+z)^3 + (1 - \Omega_{m,0})
                      ~\exp \left[ 3\int_{0}^z dz'~\frac{1+w_{DE}(z')}{1+z'}\right]~,\lb{h2z}
\ee 
with $h(z)\equiv \frac{H}{H_0}$, $w_{DE}(z)\equiv p_{DE}(z)/\rho_{DE}(z)$, $z=\frac{a_0}{a}-1$, 
and finally $\Omega_{m,0}\equiv \frac{\rho_{m,0}}{\rho_{{\rm cr},0}}$. We do not assume $w_{DE}(z)$
to be universal, i.e. independent of redshift and of initial conditions, though it has to be 
given anyway. Equality \eqref{h2z} holds for FLRW models inside GR and for many 
models beyond GR as well with appropriate definitions of the dark energy sector. 
We recall the definition $\Omega_m=\Omega_{m,0} \frac{a_0^3}{a^3} h^{-2}$
and the useful relation 
\be
w_{DE} = \frac{1}{3(1-\Omega_m)}~ \frac{d\ln \Omega_m}{d\ln a}~.  \label{wDE}
\ee
Introducing the growth function $f\equiv \frac{d \ln \delta_m}{d \ln a}$ and using 
\eqref{wDE}, equation \eqref{del} can be recast into the equivalent nonlinear first order 
equation \cite{WS98}
\be
\frac{df}{dN} + f^2 + \frac{1}{2} \left(1 - \frac{d \ln \Omega_m}{dN} \right) f = 
                              \frac{3}{2}~\Omega_m~,\lb{df}
\ee
with $N\equiv \ln a$. Clearly, for constant $p$, $f=p$ if $\delta_m\propto a^p$. 
Generically, this formalism is applied when the decaying mode is vanishingly small (see 
\cite{CFGPS19} for a more general approach). 

The following parametrization has been intensively used and investigated in the context 
of dark energy
\be
f = \Omega_m(z)^{\gamma(z)}~,\lb{Omgamma}
\ee
where $\gamma$ is dubbed growth index, though generically $\gamma(z)$ is a genuine function
which can even depend on scales for modified gravity models. 
The representation (\ref{Omgamma}) is a powerful tool in order to discriminate between 
DE models based on modified gravity theories and the $\Lambda$CDM paradigm.

In many DE models outside GR \cite{MG} the modified evolution of matter perturbations can 
be written as  
\be
{\ddot \delta_m} + 2H {\dot \delta_m} - 4\pi G_{\rm eff}\rho_m\delta_m = 0 \label{delmod}
\ee
at sufficiently small scales exceeding the effective 'Jeans' scale for cold matter but much 
smaller than the Hubble one, where $G_{\rm eff}$ is some effective gravitational coupling 
appearing in the model. For example, for effectively massless scalar-tensor models 
\cite{BEPS00}, $G_{\rm eff}$ is varying with time but it has no scale dependence, while its 
value today is equal to the usual Newton's constant $G$. Introducing for convenience the 
quantity 
\be
g\equiv \frac{G_{\rm eff}}{G}\lb{g}~,
\ee  
we easily get from \eqref{delmod} the modified version of Eq. \eqref{df}, viz.
\be
\frac{df}{dN} + f^2 + \frac{1}{2} \left(1 - \frac{d \ln \Omega_m}{dN} \right) f = 
                              \frac{3}{2}~g~\Omega_m~,\lb{dfmod}
\ee
where the same GR definition $\Omega_m = \frac{8\pi G \rho_m}{3 H^2}$ is used. 
Some subtleties can arise if the defined energy density of DE becomes negative. 
For a stricly constant growth index $\gamma$, it is straightforward to deduce from 
\eqref{dfmod} that $w_{DE}=\overline{w}$ with
\bea
\overline{w} &=& - \frac{1}{3(2\gamma -1)} ~
\frac{1 + 2\Omega_m^{\gamma} - 3 g \Omega_m^{1-\gamma}}{1-\Omega_m}~. \lb{wgcon}\\
&\equiv& - \frac{1}{3(2\gamma -1)} ~ F(\Omega_m; g, \gamma). 
\eea
The case $g=1$ reduces to GR and we will simply write
\be 
F(\Omega_m; g=1, \gamma)\equiv F(\Omega_m; \gamma)~.
\ee
Below, for a quantity $v$, $v_{\infty}$, resp. $v_{-\infty}$, will denote its (limiting) 
value for $x\to \infty$ in the DE dominated era ($\Omega_m\to 0$), resp. 
$x\to -\infty$ (generically $\Omega_m\to 1$). 
We have in particular from \eqref{wgcon} for $g=1$ (GR)
\bea
\gamma &=& \frac{3 \overline{w}_{\infty} - 1}{6 \overline{w}_{\infty}} \lb{gconstasf}\\
\gamma &=& \frac{3(1 - \overline{w}_{-\infty})}{5 - 6 \overline{w}_{-\infty}}~.\lb{gconstasp}
\eea
We assume $\overline{w}<0$ to get matter domination in the
past and DE domination in the future. Eq.\eqref{gconstasf}
requires further $\overline{w}_{\infty} < -\frac13$ in order 
to have $0< \gamma <1$, otherwise $\overline{w}_{\infty}$ becomes infinite.

It was found in \cite{PSG16} that these relations between a \emph{constant} $\g$ and the 
corresponding asymptotic values $\overline{w}_{\infty}$ and $\overline{w}_{-\infty}$ apply 
also for the \emph{dynamical} $\g$ obtained for an arbitrary but given $w_{DE}$. 
In the latter case, we obtain for $g=1$  
\bea
\gamma_{\infty} &=& \frac{3 w_{\infty} - 1}{6 w_{\infty}} \lb{gasf}\\
\gamma_{-\infty} &=& \frac{3(1 - w_{-\infty})}{5 - 6 w_{-\infty}}~,\lb{gasp}
\eea
with $w_{\infty}$, resp. $w_{-\infty}$, the asymptotic value of $w_{DE}$ in the future, resp. 
past. This nice property can actually be generalized to modified gravity models 
\cite{CFGPS19}. 
 
Taking $\Omega_m$ as the integration variable, the evolution equation for $\gamma$ 
obtained from \eqref{dfmod} using \eqref{Omgamma} yields
\begin{equation}
2\alpha \Omega_m \ln(\Omega_m) \frac{d \g}{d\Omega_m}+\alpha(2\g-1)+F(\Omega_m;g,\g)=0~, \label{dg}
\end{equation}
where we have defined 
\be
\alpha\equiv 3 w_{DE}~.
\ee
The solutions to equation \eqref{dg} on the entire $\Omega_m$ interval is the envelope of 
its tangent vectors  
$$
\left( 
\begin{array}{c}
1 \\
\frac{d \g}{d\Omega_m}
\end{array}
\right)~.
$$
All these tangent vectors define a vector field that can be written \cite{CFGPS19}
$$
\left( 
\begin{array}{c}
2\alpha \Omega_m (1-\Omega_m) \ln(\Omega_m) \\
-\alpha(2\g-1)(1-\Omega_m)-\tilde{F}(\Omega_m;\g)
\end{array}
\right)~.
$$
where we have defined 
\be
\tilde{F}(\Omega_m;g,\g)\equiv (1-\Omega_m)F(\Omega_m;g,\g)= 1 + 2\Omega_m^\g - 
                                                                  3g\Omega_m^{1-\g}~. 
\lb{tildeF}
\ee
We write the vector field in this way in order to have explicitly regular functions everywhere 
for $(\Omega_m,\g) \in [0,1] \times \mathbb{R}$.
%
%
One obtains the integral curves of this vector field (i.e. the phase portrait) by solving 
the autonomous differential system
$$
\begin{array}{l}
\frac{d\Omega_m}{ds}=2\alpha\Omega_m  (1-\Omega_m) \ln(\Omega_m) \\
\frac{d\g}{ds}=-\alpha(2\g-1)(1-\Omega_m)-\tilde{F}(\Omega_m;\g) \lb{dOgs}
\end{array}
$$
where $s \in \mathbb{R}^+$ is a dummy variable parametrizing the curves.
Clearly, the trajectories $\gamma(\Omega_m)$ are not unique. 
Only one integral curve however is finite everywhere: for $\Lambda$CDM, it is the curve 
$\gamma(\Omega_m)$ which starts (in the past) at $\g(1)\equiv \g_{-\infty} = 6/11$ and ends 
(in the future) at $\g(0)\equiv \g_{\infty} = 2/3$. It corresponds to the presence solely of 
the growing mode of Eq.\eqref{del}, or equivalently to the limit of a vanishing decaying mode. 
For cosmological constraints on DE models, one is essentially interested in that unique 
trajectory corresponding to a vanishing decaying mode. It is the only trajectory which has a 
finite initial condition $\gamma_{-\infty}$ at $\Omega_m=1$, for all other trajectories $\gamma$ 
will diverge in the past. 
However, concerning the asymptotic future ($\Omega_m \to 0$), inside GR the 
solution to Eq. \eqref{dg} gives $\g \to \gamma_{\infty}$, Eq. \eqref{gasf}, with 
($w_{\infty}<-\frac{1}{3}$)
\be
f \propto C a^{-\frac12(1-3 w_{\infty})} \to 0~. \lb{finfty}
\ee  
Indeed, we have asymptotically in the future 
\be 
\delta = \delta_{\infty}  +  {\rm const}\cdot a^{-\frac12(1-3 w_{\infty})}~,\lb{delas}
\ee
where $\delta_{\infty}$=constant corresponds to the limiting (dominant) solution of \eqref{del} 
when the last term is neglected. 
The crucial point is that the asymptotic behaviour \eqref{finfty} is identical for \emph{all} 
cases where a decaying mode of arbitrary amplitude is present, up 
to a change of the prefactor in \eqref{finfty} which depends on initial conditions and on the 
amplitude of the decaying mode with respect to the growing mode (see e.g. Figure 2 in 
\cite{CFGPS19}), including those with vanishing decaying mode. Taking into account that 
$\Omega_m \sim a^{3 w_{\infty}}$, it is straightforward to obtain from Eq.\eqref{Omgamma} that
\be
\gamma = \frac{\ln f}{\ln\Omega_m } \to \gamma_{\infty}  \lb{gamas} 
\ee
for \emph{all} curves.  
This is complementary to the results obtained in \cite{LP18}, where the growing mode for 
models beyond GR was considered.  
\section{Dynamical system approach}
In this section, we will study our equations using the dynamical system approach. 
While the introduction of the variable $\Omega_m$ is natural for a global analysis of the 
evolution of the growth index $\gamma$, we use the integration variable $N\equiv \ln a$ 
(equivalently $x=a/a_0$) for the dynamical system approach, and we obtain for $g=1$ (GR)
\begin{equation}
2 \ln(\Omega_m) \frac{d \g}{dN}+\alpha(2\g-1)(1-\Omega_m)+\tilde{F}(\Omega_m;\g)=0~.\lb{dgN}
\end{equation}
This is equivalent to the following differential system
\bea
\frac{d\Omega_m}{dN} &=& \Omega_m(1-\Omega_m) \alpha \lb{dOmN} \\
\frac{d\gamma}{dN} &=& -\frac{\alpha(2\g-1)(1-\Omega_m) + 
                         \tilde{F}(\Omega_m;\g)}{2 \ln(\Omega_m)}~.
\lb{dgN1} 
\eea
We will use these equations in order to find the critical (or stationary) points of our 
system satisfying $\frac{d\Omega_m}{dN}=0,~\frac{d\gamma}{dN}=0$.
Note that Eq.\eqref{dOmN} is independent of $\gamma$ and can therefore be integrated 
independently. When the function $\alpha(a)\equiv 3w_{DE}(a)$ is known, we can obtain 
$\alpha(\Omega_m)$ using \eqref{wDE}. 

We find readily from \eqref{dOmN} that $\frac{d\Omega_m}{dN}=0$ in the following three cases: 
$\Omega_m=0$, $\Omega_m=1$, $\alpha(\Omega_m)=0$.
The stability of a dynamical system is given by the 
Hartman-Grobman theorem which asserts that there is a certain 2$\times$2 matrix whose 
eigenvalues characterize the behavior of the system around the critical points. 
For the critical point corresponding to $\Omega_m=1$
\begin{align}
\Bigl(\Omega_m=1, \gamma = \gamma_{-\infty}\Bigr)~,\lb{cr1}
\end{align}
we find that the eigenvalues of our system are $(-2\alpha_{-\infty}, 2\alpha_{-\infty} - 5)$
and therefore the critical point is a saddle point for $\alpha_{-\infty}\leq 0$. 
For the critical point corresponding to $\Omega_m = 0$
\begin{align}
\Bigl( \Omega_m=0, \gamma = \gamma_{\infty} \Bigr )~,\lb{cr2}
\end{align}
the eigenvalues of the linearized system are $(\alpha_{\infty},0)$ and therefore we 
conclude that it is an attractor for $\alpha_{\infty}\leq 0$. Notice that the zero eigenvalue 
does not point to any stability or instability, but a simple centre manifold analysis 
allows us to conclude about the stability of the critical point. 
To study the structure of the phase space at infinity, we define $u=1/\gamma$. 
We obtain that $u=0$ ($\gamma=\pm \infty$) is also a critical point and it is easy 
to show that $u=0$ is a repeller.
These results of the dynamical system analysis confirm the asymptotic properties found 
analytically and numerically in \cite{CFGPS19} and summarized in Section 2.

\begin{figure}
\begin{centering}
\includegraphics[scale=.5]{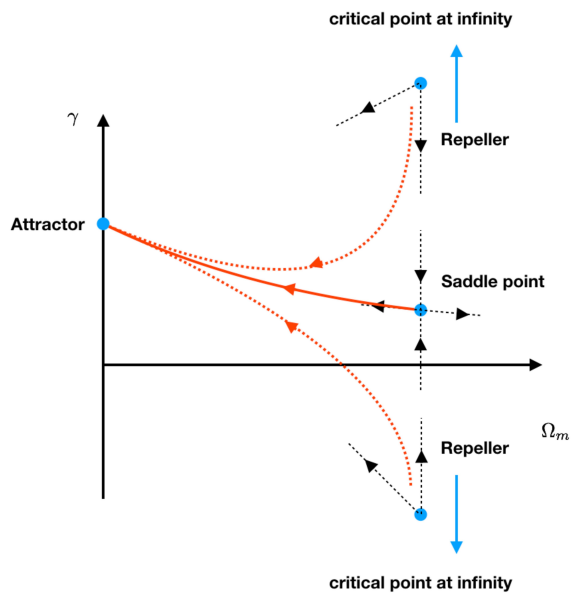}
\par\end{centering}
\caption{The phase portrait corresponding to the system of Eqs.\eqref{dOmN},\eqref{dgN1}. We 
have an attractor at $(\Omega_m=0,\gamma_{\infty})$, a saddle point at 
$(\Omega_m=1,\gamma_{-\infty})$ and the infinity which is a repeller. The red line 
corresponds to the trajectory connecting the 2 critical points (known as an heteroclinic 
orbit) while the red dotted lines connects infinity to the attractor. This phase portrait 
illustrates nicely the asymptotic properties of the trajectories presented in Section 2.}  
\label{FigStructure}
\end{figure}

The remaining critical points correspond to $\alpha(\Omega_m)=0$ and $\tilde{F}=0$.
Indeed, various critical points can exist if $\alpha(\Omega_m)$ has different zeroes. 
These critical points can have a richer structure. The eigenvalues associated to this 
system are $(-2\Omega_m^\gamma-1/2,\Omega_m(1-\Omega_m)\alpha'(\Omega_m))$. 
If $\frac{d\alpha}{d\Omega_m}\equiv \alpha'(\Omega_m)<0$, the critical point is an attractor, 
if $\alpha'(\Omega_m)>0$, the critical point is a saddle point. 
In particular for constant $\gamma$, these critical points correspond 
to the family of tracking DE solutions for $\Omega_m < 1$ and $\tilde{F}=0$ found 
in \cite{PSG16}.
As $\frac{d\overline{w}}{d\Omega_m}>0$ in this case, our calculations confirm that these 
critical points correspond to saddle points.
In the simpler (and generic) case where $\alpha$ has no zeroes ($\alpha<0$), we can sketch 
the evolution of the system in the phase space $(\Omega_m,\gamma)$.
(Fig.(\ref{FigStructure})).
We see from Fig.(\ref{FigStructure}), the existence of a special orbit, known as an 
heteroclinic orbit which connects the 2 critical points. Because it follows the 
repelling direction of the saddle point, it is easy to find from the eigenvector of the 
linearized system the behavior of this orbit around the critical point at $\Omega_m=1$ 
and one obtains
\begin{align}
\label{eq:heteroclinic}
\gamma = \gamma_{-\infty} + \frac{(\alpha_{-\infty} - 2)(\alpha_{-\infty} -3) +
2 \alpha'_{-\infty}(2\alpha_{-\infty}-5)} {2(5-4 \alpha_{-\infty})(5-2\alpha_{-\infty})^2} 
                              (1-\Omega_m)+\mathcal{O}\Bigl((1-\Omega_m)^2\Bigr)~.
\end{align}
This line is the asymptote of the heteroclinic orbit. 
Note that \eqref{eq:heteroclinic} generalizes the result given in \cite{PSG16} for 
constant $\alpha$. One checks easily that $\Lambda$CDM satisfies indeed 
\eqref{eq:heteroclinic}.

Because we consider a dynamical system (system of first order differential equations), the 
trajectories (orbits) in phase space cannot intersect. But of course other curves which 
are not orbits of the system can intersect these orbits, e.g. we can consider the curve 
for which $\frac{d\gamma}{dN}=0$ everywhere. 
For $\alpha < 0$, it satisfies $\frac{d\gamma}{d\Omega_m}=0$ for $0 < \Omega_m < 1$ and goes 
through the endpoints $\gamma_{-\infty}$, resp. $\gamma_{\infty}$, at $\Omega_m = 1$, resp. 
$\Omega_m = 0$. from eqs.\eqref{cr1},\eqref{cr2}.So it corresponds to the curve dubbed 
$\Gamma(\Omega_m)$ in \cite{PSG16}. 
It satisfies $\overline{w}(\Omega_m,~\Gamma(\Omega_m)=w_{DE}(\Omega_m)$ and we have indeed 
$\Gamma(1)=\gamma_{-\infty}$ and $\Gamma(0)=\gamma_{\infty}$. 
For arbitrary $w_{DE}$, $\gamma$ is not constant and hence $\Gamma(\Omega_m)$ is not 
constant either. 
As $\Gamma$ satisfies by construction $\frac{d\gamma}{dN}=0$ and critical points are defined 
by $\frac{d\gamma}{dN}=0,\frac{d\Omega_m}{dN}=0)$, $\Gamma$ must intersect the critical points, 
but of course it can also intersect orbits at points which are not critical points.
 
We can ask if it is above or under the heteroclinic orbit that we previously defined because 
they start and end at the same points. A global analysis is impossible, but we can at least 
analyze the behavior around $\Omega_m=1$. 
We have already found the tangent to the heteroclinic orbit (see Eq.\ref{eq:heteroclinic}). 
We can also calculate the tangent to $\Gamma$ and we find around $\Omega_m = 1$
\begin{align}
\Gamma = \gamma_{-\infty} + \frac{(\alpha_{-\infty} - 2)(\alpha_{-\infty} - 3) + 
2\alpha'_{-\infty}(2\alpha_{-\infty}-5)}
{2(5-2\alpha_{-\infty})^3}(1-\Omega_m) + \mathcal{O}\Bigl((1-\Omega_m)^2\Bigr)~.\lb{TanGam}
\end{align}
Therefore $\Gamma$ lies above the heteroclinic orbit iff
\begin{align}
\alpha'_{-\infty} < \frac{(\alpha_{-\infty}-2)(\alpha_{-\infty}-3)}{10-4 \alpha_{-\infty}}~.\lb{Gamgr}
\end{align}
One checks easily that $\Lambda$CDM satisfies \eqref{Gamgr}.

These results can be easily generalized to modified gravity for which the system becomes 
\begin{align}
\frac{d\Omega_m}{dN} &=\alpha  (1-\Omega_{m}) \Omega_{m}\\
\frac{d\gamma}{dN} &= -\frac{\alpha  (2 \gamma -1) (1-\Omega_{m}) + 
\tilde{F}(\Omega_m; g, \gamma)}{2\ln (\Omega_{m})}
\end{align}
We recover the same critical points as in GR if $g=1$.
Note that $g_{-\infty}=1$ and $\left(\frac{dg}{dN}\right)_{-\infty}=0$ in order to avoid that 
$w_{-\infty}$ becomes singular \cite{PSG16}. 
The coordinate of the critical point at $\Omega_m=1$ changes into
\begin{align}
\Bigl(\Omega_m = 1,\gamma = \frac{\alpha_{-\infty} - 3 - 
                            3g'_{-\infty}}{2 \alpha_{-\infty} - 5}\Bigr)~.
\lb{crg}
\end{align}
As expected, the expression for $\gamma$ in Eq.\eqref{crg} corresponds to the only finite 
value in the asymptotic past found earlier \cite{CFGPS19}. 
Finally, we can also find the condition for which the curve $\Gamma$ starts at $\Omega_m=1$ 
with an inclination larger than that of the heteroclinic orbit, viz.
\bea
\alpha'_{-\infty} &<& \frac{(\alpha_{-\infty}-2)(\alpha_{-\infty}-3)}{10-4 \alpha_{-\infty}} + 
\frac{3}{2}g''_{-\infty}(2\alpha_{-\infty}-5) \nonumber \\
+ &3& g' _ {-\infty}\frac { 33 - 28\alpha_ {-\infty} + 6\alpha_ {-\infty}^2 - 27 g' _ {-\infty} + 
  12\alpha_ {-\infty} g' _ {-\infty} - 6 (5 - 2\alpha_ {-\infty})^2 g'' _ {-\infty} }
{2 (2\alpha_{-\infty} - 5) (6 g'_ {-\infty} + 1)}
~.\lb{Gammodgr}
\eea 

While the DGP model is excluded by observations, it is still a
very popular modified gravity DE toy model because of its connections with
quantum field and string theory. Its study is surprisingly simple in our formalism 
because both $g^{DGP}$ and $w^{DGP}$ can be expressed in function of $\Omega_m$.
When we apply \eqref{Gammodgr} to the Dvali-Gabadadze-Porrati (DGP) model \cite{DGP00}
([$g^{DGP}]'_{-\infty} = \frac13$), it is found that the inequality \eqref{Gammodgr} is satisfied. 
Hence the heteroclinic orbit in the DGP model is a decreasing function 
of $\Omega_m$ in the neighbourhood of $\Omega_m=1$. This contrasts with the general shape of 
the heteroclinic orbit in the DGP model: it is an increasing function of $\Omega_m$ except 
for $\Omega_m\lesssim 10^{-3}$ and $\Omega_m\gtrsim 0.9$, the latter decrease (in the 
asymptotic past) is very tiny as compared to the sharp decrease in the asymptotic future 
\cite{CFGPS19}.   
\section{Presence of an unclustered dust-like component}
We consider yet another case inside GR where the growth index $\g(\Omega_m)$ is not 
monotonically decreasing in contrast to $\Lambda$CDM. Let us note first that in the 
particular case where $\Omega_m$ is constant, we readily get from \eqref{df} 
\be
f' + f^2 + \frac12 f - \frac32 C = 0~, \lb{dfC}
\ee
and we set $\Omega_m = C$ to emphasize the constancy of $\Omega_m$. 
Equation \eqref{dfC} has two constant solutions 
\be
p_1 = -\frac14 + \frac14 \sqrt{1 + 24C}~,~~~~~~~~~~~~~~
                          p_2 = -\frac14 - \frac14 \sqrt{1 + 24C}~. \lb{p12C}
\ee
For $C>0$, we have necessarily $p_1>0$ and $p_2<0$. In other words, there are 
two genuinely growing and decaying modes for $\delta_m$. 
When $C=1$ we recover the standard results in an Einstein-de Sitter universe.

An interesting situation, still in the absence of DE, arises when dust-like matter has 
some (small) relative fraction $\Omega_x$ which does not cluster and only usual matter 
denoted by $\Omega_m$ does, with $\Omega_m + \Omega_x = 1$. 
Phenomenologically, this unclustered component could be ultra-light dark matter, or 
DE tracking matter \emph{exactly} though later we will consider the presence of a DE component 
different from the unclustered dust-like matter component. It can also represent a light 
relativistic species like massive neutrinos once they become non-relativistic. 

Then Eq.\eqref{dfC} is obtained with $C=\Omega_m<1$. 
Let us consider for concreteness the situation with  $\Omega_x\ll 1$.
From \eqref{p12C} the growing mode scales $\propto a^{p_1}$ with 
\be
p_1\approx 1-\frac35 \Omega_x\approx \Omega_m^{\frac35}~. \lb{p1nu}
\ee   
The last term in \eqref{p1nu} makes contact with the growth index $\gamma$. In the case under 
consideration, both $\Omega_m$ and $f=p_1>0$ are constant, hence $\gamma$ is constant, too, 
and from \eqref{p1nu} it is close to $\frac35$ (see the nice discussion in \cite{T05}).

In \cite{PSG16}, a family of solutions with constant $\gamma > \frac35$ was found corresponding 
to the roots of $F(\Omega_m;\gamma)$ for $\Omega_m<1$ with 
$\overline{w}=0$ so that $\Omega_m$ remains constant. This corresponds to our system 
with $\overline{w} = w_x$. For $\Omega_x\ll 1$ it was found \cite{PSG16} 
\be
p_1 = \Omega_m^{\frac35 \left(1 + \frac{\Omega_x}{25} \right)}~,  \lb{p1nub}
\ee 
when we expand $\gamma$ up to first order in $\Omega_x$. We see that \eqref{p1nub} refines 
the result \eqref{p1nu} (see also \cite{LC07}, \cite{WS98}). 

We now extend these results to a universe where the expansion is driven also by an additional 
non-tracking (genuine) dark energy component so that $\Omega_m$ is no longer constant. 
Using for convenience the variable $\ot$ instead of $\Omega_m$, the growth index satisfies 
the equation 
\begin{equation}
\left[ 2 ~\ot \ln\left((1-\e)\ot \right) \frac{d \g^{\e}}{d\ot} 
    + (2\g^{\e}-1) \right]~\alpha (1-\ot)  +  \tilde{F}((1-\e)\ot;\g^{\e})= 0~, \lb{dgam}
\end{equation}
with 
\be
\Omega_{DE} = 1 - \Omega_m^{\rm tot}~,~~~~~~~~\Omega_m =  (1 - \varepsilon) \ot ~, 
~~~~~~~~~~\Omega_x=\e~\ot~,\lb{OmDE}
\ee
obviously satisfying 
\be
\ot + \Omega_{DE}\equiv \Omega_m + \Omega_x + \Omega_{DE} = 1~.\lb{Omtot}
\ee
We have noted $\g^{\e}(\ot)$ the solution of Eq.\eqref{dgam} for $\e>0$.
In the asymptotic past, $\Omega_{DE}\to 0$, \eqref{dgam} becomes 
\be
- 6 ~w_{-\infty} \ln (1-\varepsilon) \frac{d\gamma^{\e}}{d\ln (1-\Omega_m^{\rm tot})} 
                  + \tilde{F}(1-\e;\g^{\e})=0~.      \lb{dgampast}
\ee
It is seen from \eqref{dgampast} that any finite solution $\gamma^{\e}$ of \eqref{dgam} 
must tend in the past to the root of $\tilde{F}(1-\e;\g)$, viz. 
\be
\tilde{F}(1-\e;\g_{-\infty}^{\e}) = 0~, \lb{root}
\ee
with $\g_{-\infty}^{\e}\equiv \g^{\e}(\ot \to 1)$.
Considering the change of variable, $X=(1-\varepsilon)^{\g_{-\infty}^{\e}}$, 
\eqref{root} transforms into $2X^2+X-3(1-\varepsilon)=0$,
whose solutions are $X_\pm=\frac{-1\pm \sqrt{25-24\varepsilon}}{4}$.
Considering only the positive root, we get
\begin{align}
    \g_{-\infty}^{\e}=\frac{\ln\Bigl(\frac{-1+\sqrt{25-24\varepsilon}}{4}\Bigr)}
                                                        {\ln(1-\varepsilon)}\lb{ginfeps}
\end{align}
%
%
Expanding this expression in series near $\e=0$ leads to
\be\lb{serieg}
\g_{-\infty}^{\e}\equiv \g^{\e}(\ot\to 1) = \frac35 + \frac{3}{125}\e+\frac{97}{6250}\e^2 +
 \frac{737}{62500}\e^3+{\cal O}(\e^4)
\ee
We recover the first two terms, the root of $\tilde{F}(1-\e;\g)$ to first order in 
$\e$ \cite{PSG16} mentioned above, Eq.\eqref{p1nub}.
Expression \eqref{serieg} extends these earlier calculations to third order in $\e$.

We note the intriguing property that $\gamma_{-\infty}^{\e}$ does not depend on any nonzero 
$w_{-\infty}$. 
In order to understand this, we consider the corresponding vector field 
$\mathcal{F}[\ot,\g^{\e};\e]$ \cite{CFGPS19}
\begin{equation}\label{vecfield2}
\mathcal{F}[\ot,\g^{\e};\e]=
\left( 
\begin{array}{c}
2\alpha \ot (1-\ot) \ln((1-\e)\ot) \\
-\alpha(2\g^{\e}-1)(1-\ot)-\tilde{F}((1-\e)\ot;\g^{\e})
\end{array}
\right)~,
\end{equation}
tangent to the solutions $\g^{\e}$ and we look for its streamlines (see \eqref{dOgs}).
For $\ot \simeq 1$ and $\e \simeq 0$, we can write the vector field (\ref{vecfield2}) 
to leading order 
\begin{equation}
\mathcal{F}[\ot,\g^{\e};\e] \simeq
\left( 
\begin{array}{c}
(\ot - 1)~\e  ~6 w_{-\infty}  \\
(\ot - 1)\left[(6 w_{-\infty} - 5)\g^{\e} 
                          - 3 w_{-\infty} + 3 \right] + \e( 3 - 5\g^{\e} )\lb{vecexp}
\end{array}
\right)~.
\end{equation}
It is seen that the leading order of the upper component $\left(\frac{d \ot}{ds}\right)$ 
is of order $(\ot-1)\times \e$ in the small parameters ($\ot-1$) and $\e$. 
In the lower component $\left(\frac{d \g}{ds}\right)$, we have neglected all higher order 
terms. 
For $\e = 0$ ($\Omega_x = 0$), in the neighbourhood of $\ot= 1$, we obtain to leading order 
in $\ot (= \Omega_m)$
\begin{equation}
\mathcal{F}[\ot,\g;0] \simeq 
\left( 
\begin{array}{c}
0 \\
(\ot - 1)\left[(6 w_{-\infty} - 5)\g_{-\infty} - 3 w_{-\infty} + 3 \right]\lb{vecexp2}
\end{array}
\right)~,
\end{equation}
with 
\be
\g_{-\infty}\equiv \g(\ot \to 1,~\e=0)~.
\ee
To avoid that $\frac{d\g}{d\ot}\Big|_{\e=0}$ diverges in the neighborhood of $\ot = 1$, 
the lower component of $\mathcal{F}[\ot\simeq 1,\g;\e=0]$ must vanish too and 
hence we get 
\be\lb{ginf}
\g_{-\infty}=\frac{3 w_{-\infty} -3}{6 w_{-\infty} - 5}~, 
\ee
so we recover the (expected) result, Eq.\eqref{gasp}. 
On the other hand, for $\e >0$ fixed and however small, from \eqref{vecexp} the limit 
$\ot \to 1$ gives to leading order in the small parameter $\e$
\begin{equation}
\mathcal{F}[\ot \to 1,\g^{\e};\e] \simeq
\left( 
\begin{array}{c}
0 \\
\e~\left(- 5\g_{-\infty}^{\e} + 3 \right)
\end{array}
\right)~.\lb{vfieldeps}
\end{equation}
We obtain now $\frac35$ to lowest order in $\e$, viz. 
\be
\g^{\e}_{-\infty} = 3/5 + {\cal O}(\e)~, \lb{pse}
\ee
in agreement with \eqref{p1nub} or \eqref{serieg}.
Actually, if we take the limit $\ot \to 1$ in \eqref{vecfield2},
without expanding in $\e$, we obtain
\begin{equation}
\mathcal{F}[\ot \to 1,\g^{\e}_{-\infty};\e] =
\left( 
\begin{array}{c}
0 \\
-\tilde{F}\left( 1-\e;\g_{-\infty}^{\e} \right)
\end{array}
\right)~, \lb{vfield1full}
\end{equation}
showing again that $\g_{-\infty}^{\e}$ must be a root of $\tilde{F}(1-\e;\g)$, and 
the value $\frac35$ obtained from \eqref{vfieldeps} is just the 
lowest order of the expansion of $\g_{-\infty}^{\e}$ in powers of $\e$.

Interestingly, there is another situation where an identical result appears \cite{PSG16}. 
Let us assume that we have a two-component system ($\e=0$) with 
$\ot=\Omega_m\to 1-\delta,~\Omega_{DE}\to \delta$. 
This is possible only if DE behaves asymptotically like dust, $w_{-\infty}=0$. If we take 
$\delta>0$, we have from \eqref{vecfield2} in analogy with \eqref{vfield1full} 
\begin{equation}
\mathcal{F}\left[1-\delta,\g_{-\infty}^{\delta};\delta \right] \simeq
\left( 
\begin{array}{c}
0 \\
-\tilde{F}\left( 1-\delta;\g_{-\infty}^{\delta} \right)
\end{array}
\right)~, \lb{vfield1-dfull}
\end{equation}
which is just $\eqref{vfield1full}$ with $\e$ replaced by $\delta$. So in this case, the 
small parameter that goes to zero is $w$ instead of $1-\ot$ previously.  

To summarize, the system with a small amount of unclustered dustlike component is not 
continuous in the variables $(\ot,\e)$ at $\ot=1,~\e=0$ and 
taking the limit is affected by the order in which it is taken, viz
\be
\lim_{\e \to 0} \gamma^{\e}_{-\infty}\equiv \lim_{\e \to 0} \gamma^{\e}(\ot \to 1) 
           \ne \lim_{\Omega_m \to 1} \gamma(\Omega_m) \equiv \gamma_{-\infty} ~.\lb{notcon}
\ee
This explains why it is possible that $\gamma_{-\infty}^{\e}$ does not depend on $w_{-\infty}$. 
We recover consistently from \eqref{vecfield2} that for 
$\e=0,~\ot=\Omega_m\to 1-\delta \ne 1,~w_{-\infty}=0$ (tracking DE), 
roots of $\tilde{F}\left(1-\delta;\g \right)$ yield the tracking DE solutions with a constant 
growth index $\g$ found in \cite{PSG16}. 

\begin{figure}[h!]
   \begin{center}
   \includegraphics[height=6cm]{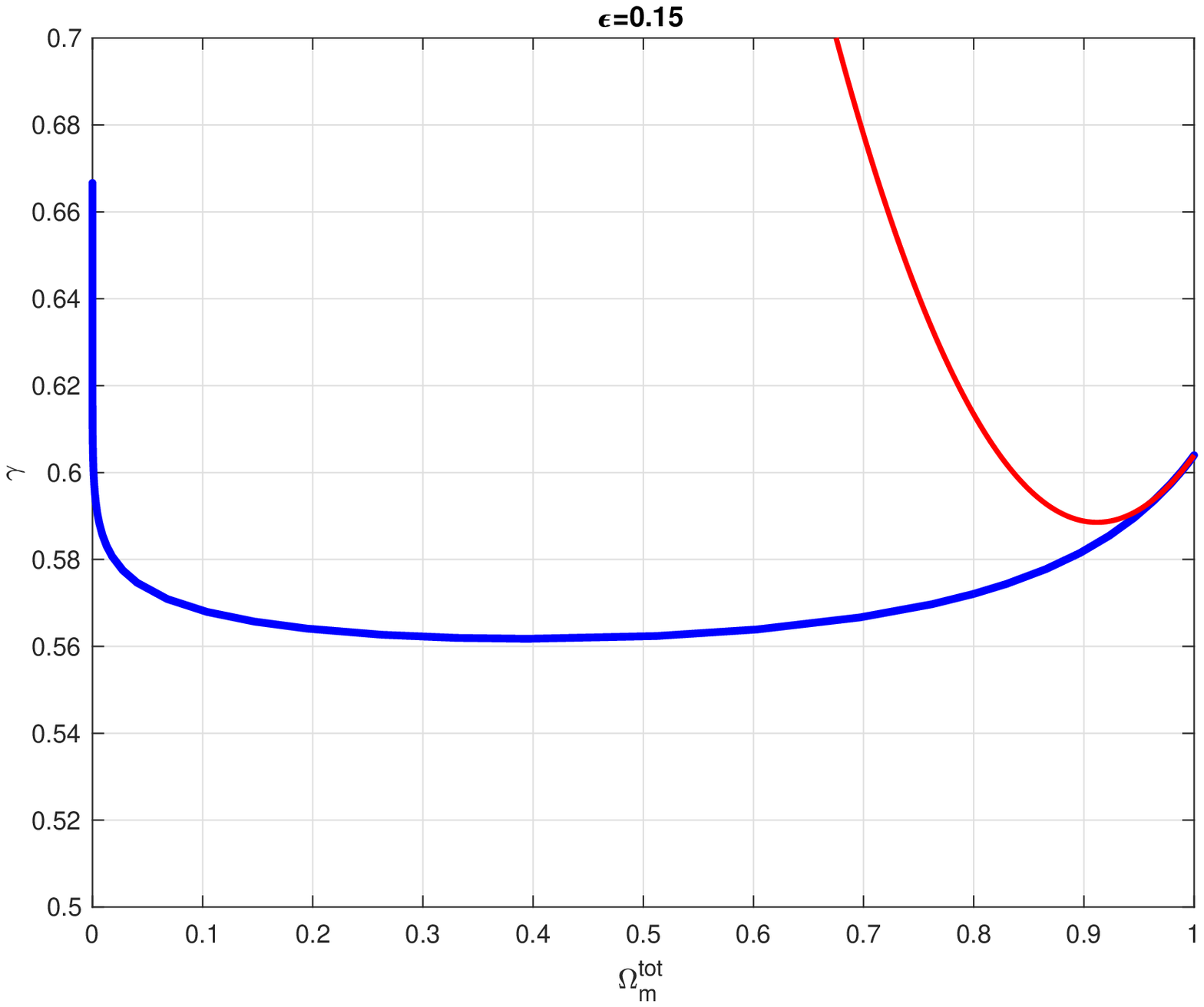}~~
   \includegraphics[height=6cm]{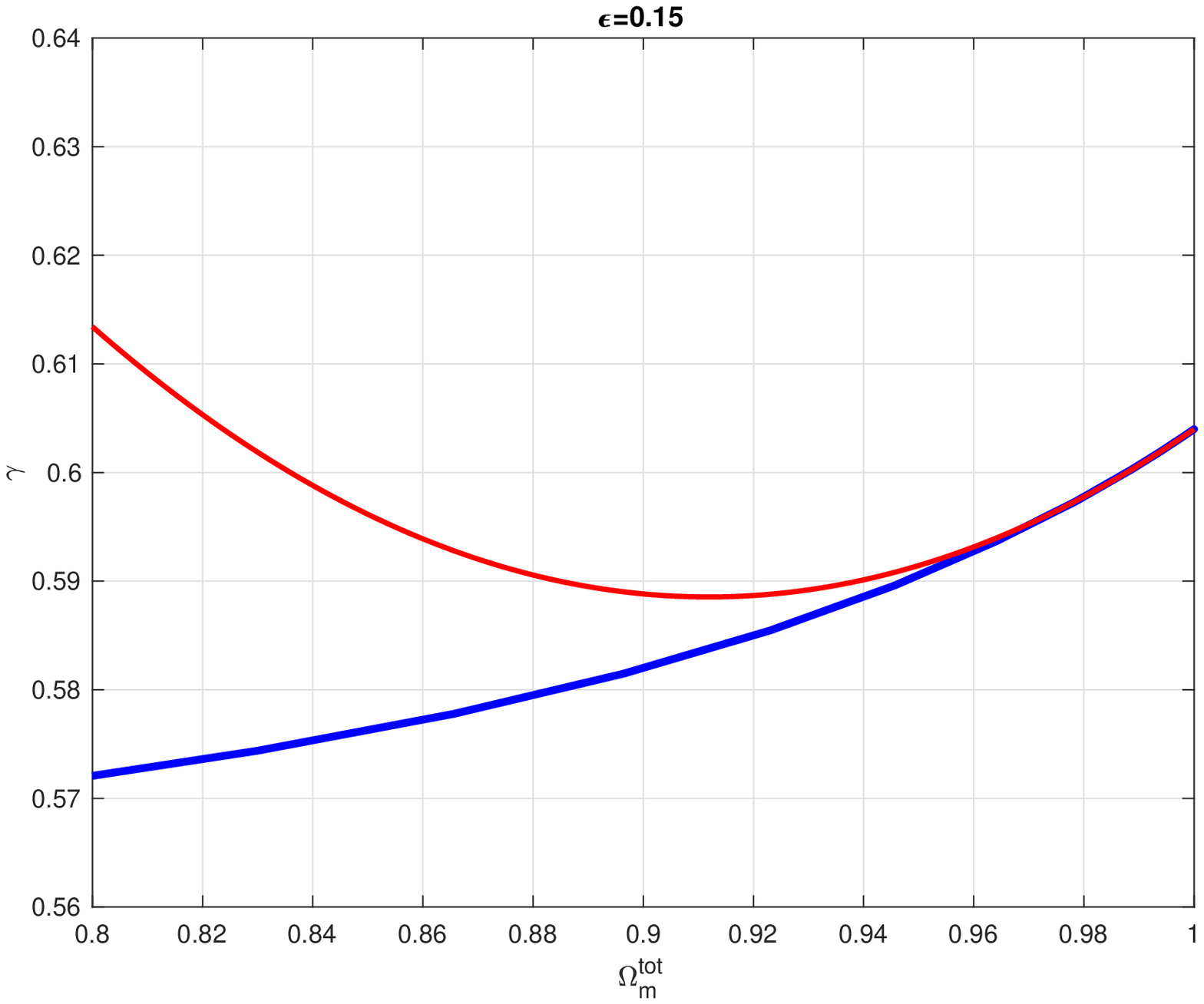}
   \caption{Left panel: The blue curve shows the reconstruction of $\gamma(\ot,\e)$ for 
   $\varepsilon=0.15$. The red curve is the second order approximation given by 
   (\ref{approx2}). 
   We see that they match nicely for $\ot\simeq 1$. Right panel shows a zoom for 
   $\ot\simeq 1$}\lb{figmatch}
   \end{center}
\end{figure}
%
%
After some calculations, (see Appendix, Eqs.\eqref{dg2}, \eqref{ddg2}), we obtain for 
$\alpha=-3$ ($w_{DE}=-1$) 
\bea
\dg\equiv \frac{d\g^{\e}}{d\ot}\Bigg|_{-\infty} =& \frac{3}{55\e}- \frac{399}{30250} - 
                                                    \frac{10161\e}{16637500} \\
\ddg\equiv \frac{d^2\g^{\e}}{d(\ot)^2}\Bigg|_{-\infty} =& \frac{6}{55 \e^2}-\frac{31083}{257125 \e}
+\frac{12960073}{1202059375}.
\end{eqnarray}
These results are illustrated with Figure \eqref{figmatch} where we can see that the 
approximation \eqref{approx2} gives an excellent match to the solution $\g^{\e}(\ot)$ for 
$\e=0.15,~\ot\simeq 1$. 
Note that we have obtained a double expansion with respect to both $\ot$ and $\e$ 
(neglecting all terms ${\cal O}(\e)$)
\bea
\g^{\e}(\ot) &\simeq& \frac{3}{5} + \left[ \frac{3}{55\e}- \frac{399}{30250}\right](\ot-1)
                                                                             \nonumber\\
&+& \frac{1}{2}\left[ \frac{6}{55 \e^2}-\frac{31083}{257125 \e}
+\frac{12960073}{1202059375} \right](\ot-1)^2 + ... \lb{dexp}
\eea
This suggests that, as a function of $\e$, $\gamma^{\e}$ has an essential singularity 
(i.e. a pole of infinite order) at $\e=0$. These expressions given above show explicitly the
discontinuity expressed by \eqref{notcon}. 
For $\e>0$, $\gamma$ tends in the asymptotic past to $\g^{\e}_{-\infty}$, which has been  
calculated here up to terms ${\cal O}(\e^4)$, (eq.\eqref{serieg}).
For $\e=0$ on the other hand, $\gamma$ tends to $\g_{-\infty}$ (eq.\eqref{gasp}). 
While $\g_{-\infty}$ depends on $w_{-\infty}$, $\g^{\e}_{-\infty}$ does not and is in this sense 
universal. Clearly it could not be possible to recover $\g_{-\infty}$ from 
$\g^{\e}_{-\infty}$ by taking the limit $\e\to 0$. This is what the equations given above show: 
while $\g^{\e}_{-\infty}$ is consistently obtained from \eqref{dexp} at $\ot=1$, the limit  
$\e\to 0$ does not even exist in the neighbourhood $\ot\simeq 1$. Note that at $\ot=1$, the 
limit $\e\to 0$ gives $\frac35$. 

We will see now other situations were the value $\frac35$ appears.
Until now we were interested in an unclustered component which behaves like dust, hence 
$\e=$ constant. For such a system we see from \eqref{dexp} that 
$\g^{\e}(\ot \to 1)\to \frac{3}{5}$ 
up to terms ${\cal O}(\e)$. 
The same limit is obtained if the unclustered component instead of behaving like dust tends 
to such a behaviour in the past, in other words if it is a tracking component in the past with
$\e$ tending to some constant value. Finally we note that our results hold for $\e<0$, see 
e.g. \cite{BGPS15}. 
 
It is also interesting to consider an unclustered component with $w_{DE}<w_{\rm uncl}<0$. 
Specializing to $w_{DE}=-1$, taking only the leading order term in $\e$ at each order of 
the expansion \eqref{dexp}, we obtain 
\bea
    \gamma^{\e}(\Omega_{DE}) &=& \frac{3}{5} + \frac{3}{55} \frac{\Omega_{DE}}{\varepsilon} +  
              \frac{3}{55} \Bigl(\frac{\Omega_{DE}}{\varepsilon}\Bigr)^2
+ \frac{3}{55} \Bigl(\frac{\Omega_{DE}}{\varepsilon}\Bigr)^3 + \cdots\\
 &=& \frac{3}{5}\Bigl[1 + 
 \frac{1}{11}\sum_{k=1}^\infty \Bigl(\frac{\Omega_{DE}}{\varepsilon}\Bigr)^k\Bigr]+\cdots\\
 &=& \frac{3}{5}\Bigl[1+\frac{1}{11}\Bigl(-1 + 
       \sum_{k=0}^\infty \Bigl(\frac{\Omega_{DE}}{\varepsilon}\Bigr)^k\Bigr)\Bigr]+\cdots\\ 
 &=& \frac{6}{11}+\frac{3}{55}\sum_{k=0}^\infty \Bigl(\frac{\Omega_{DE}}{\varepsilon}\Bigr)^k + 
                                \cdots  \label{eq:expansion}
\eea
For our system, $\frac{\Omega_{DE}}{\varepsilon}\to 0$ so the sum is well-defined and 
\be
    \gamma^{\e}(\Omega_{DE}) = \frac{6}{11} + 
             \frac{3}{55}\frac{1}{1 -\frac{\Omega_{DE}}{\varepsilon} }+\cdots
\ee
For $\frac{\Omega_{DE}}{\varepsilon}\to 0$ we obtain again
\be
    \gamma^{\e}(\Omega_{DE}\to 0) \to \frac{6}{11}+\frac{3}{55}=\frac{3}{5}~.
\ee

\section{Summary and conclusion}
The growth index $\gamma$ is a interesting tool for the study of the evolution of matter 
perturbations on cosmic scales in various cosmological models (see e.g. \cite{gamma} for 
its use in different contexts). Though it was introduced in order to characterize the 
influence of 
a non-vanishing spatial curvature on the growth of matter perturbations, interest for its 
study was revived in the context of DE models. Indeed, the growth index is a 
particularly efficient tool for the assessment of DE models in modified gravity.
    
We are interested in the global dynamics of $\gamma$ from the asymptotic past to the 
asymptotic future. Though only a restricted interval of redshifts is relevant for observations, 
a global analysis yields a deeper insight \cite{CFGPS19}.
Using the dynamical system approach we have found all critical points of the system. 
That unique trajectory for which the growth index remains finite from the asymptotic future 
to the asymptotic past is identified as the heteroclinic orbit connecting the critical points
$\left(\Omega_m=0,~\gamma_{\infty}\right)$ in the asymptotic future and 
$\left(\Omega_m=1,~\gamma_{-\infty}\right)$ in the asymptotic past. The critical point 
$\left(\Omega_m=0,~\gamma_{\infty}\right)$ is an attractor while the critical point 
$\left(\Omega_m=1,~\gamma_{-\infty}\right)$ is a saddle point. These results confirm our 
earlier findings \cite{CFGPS19}. We recall that this unique trajectory corresponds to a
vanishing decaying mode. As an additional result, we have refined our earlier results regarding 
the behaviour of $\gamma(\Omega_m)$ in the DGP model and we find its very tiny decrease in the 
past, while it is essentially an increasing function except in the asymptotic future 
($\Omega_m\lesssim 10^{-3}$). 
 
We have considered a system consisting of DE with an effective equation of state having 
arbitrary dependence on redshift and partially clustered dust-like matter with some (small) 
component of the latter remaining smooth at all scales, and investigated the growth of 
perturbations in it at scales exceeding the Jeans (or free streaming) length of 
gravitationally clustered matter (but much less than the Hubble scale). 
We have shown both analytically and numerically that $\g_{-\infty}^{\e}$ is the root of 
$\tilde{F}(1-\e;\g)$ for $\Omega_m\to 1-\e<1$ and we have calculated $\g_{-\infty}^{\e}$ to 
third order in the small parameter $\e$. 
Interestingly $\g_{-\infty}^{\e}$ does not depend on $w_{DE}$ which is possible because, as we 
have shown $\lim_{\e\to 0} \g_{-\infty}^{\e}\ne \g_{-\infty}$ where the last quantity corresponds to 
(usual) clustered dust and depends of course on $w_{DE}$. The quantity $\g_{-\infty}^{\e}$ was 
found earlier to correspond to the constant growth index corresponding to tracking DE
in the matter era with $\Omega_{DE}\to \e$. We find further that 
$\frac{d\g_{-\infty}^{\e}}{d\Omega_m}\sim \frac{1}{\e^2}$ for $\e\simeq 0$ suggesting that 
$\g_{-\infty}^{\e}$ has an essential singularity at $\e=0$. 
The results presented in this work show that besides its use for the assessment of DE models, 
the growth index $\gamma$ has also interesting mathematical properties 
as we have seen when dealing with partly unclustered dust-like matter. 

\section*{Acknowledgements}

R.G. is supported by Fondecyt project No 1171384. A.A.S. was partially supported by the Project 
KP19-261 of the Presidium of the Russian Academy of Sciences "Physics of hadrons, leptons, 
Higgs boson and dark matter particles" and by the project number 0033-2019-0005 of the Russian 
Ministry of Science and Higher Education.

\section*{Appendix}
In order to evaluate the derivative of $\g^{\e}(\ot)$ with respect to $\ot$ at $\ot=1$, 
let us assume $\g^{\e}(\ot)$ is analytic with respect to $\ot$ and use an expansion at 
second order in $(\ot-1)$, viz.
\begin{equation}\label{approx2}
\g^{\e}(\ot)= \g_{-\infty}^{\e} + \frac{d\g^{\e}}{d\ot}\Bigg|_{-\infty} (\ot-1) + 
                 \frac{1}{2}\frac{d^2\g^{\e}}{d(\ot)^2}\Bigg|_{-\infty} (\ot-1)^2 
                                          + {\cal O}\left((\ot-1)^3 \right)~.
\end{equation}
For simplicity, we denote 
\be
\dg\equiv \frac{d\g^{\e}}{d\ot}\Bigg|_{-\infty} ,\,\,\,\,\,\, 
         \ddg\equiv \frac{d^2\g^{\e}}{d(\ot)^2}\Bigg|_{-\infty}~.
\ee
In the neighbourhood of $\ot=1$, the derivative $\frac{d\g^{\e}}{d\ot}$ has therefore 
the following expansion 
$$
\frac{d\g^{\e}}{d\ot}=\dg+ \ddg (\ot-1) + {\cal O}\left((\ot-1)^2 \right)~.
$$
We will derive in this Appendix an expression for $\dg$ and $\ddg$ as functions of $\e$.
Let us use this expansion in the vector field $\mathcal{F}[\ot,\g^{\e};\e]$, 
\eqref{vecfield2}, and compute the ratio of the components, which then gives the 
derivative $\frac{d\g^{\e}}{d\ot}\Big|_{-\infty}$. We will assume here that $\alpha$ is constant. 
The upper component of $\mathcal{F}[\ot,\g^{\e};\e]$, \eqref{vecfield2}, is
\begin{equation}
2\alpha \ot (1-\ot) \ln\left( (1-\e)\ot \right) \simeq - 2\alpha \left[ \ln(1 - \e) (\ot-1) + 
                                  (\ln(1 - \e) + 1)(\ot-1)^2 \right]~, 
\end{equation}
while the lower component is
\begin{eqnarray}
-\alpha(2\g^{\e}-1)(1-\ot) &-& \tilde{F}\left( (1-\e)\ot;\g_{-\infty}^{\e} 
             + \dg~(\ot-1)+1/2 \ddg~(\ot-1)^2 \right)
\nonumber \\
&\equiv& f_0(\e)+f_1(\e) (\ot-1) + f_2(\e) (\ot-1)^2 + \hdots
\end{eqnarray}
Considering $\Omega_m=1$, it is trivial to find 
\begin{align}
  f_0(\e) =  -\tilde{F}\left( (1-\e);\g_{-\infty}^{\e}\right) = 0~,\lb{f0}
\end{align}
where the last equality comes from Eq.(\ref{root}). 
The derivative $\frac{d\g}{d\ot}$ in the neigbourhood of $\ot=1$ is therefore given by
\begin{equation}\lb{derg}
\frac{d\g}{d\ot} = \frac{f_1(\e) + f_2(\e) (\ot-1) +\hdots}
             {- 2\alpha \ln (1 - \e) - 2\alpha(\ln(1 - \e) + 1)(\ot-1)}~.
\end{equation}
We can proceed to the identification
\bea
\dg &=& \frac{f_1(\e)}{-2\alpha \ln(1-\e)}~, \\
\ddg &=& \frac{f_2(\e)}{- 2\alpha \ln (1 - \e)} - 
            \frac{f_1(\e)(- 2\alpha( \ln (1 - \e) + 1 ) )}{(- 2\alpha \ln (1 - \e))^2}~.
\eea
We have further
\bea
    f_1(\e) &=& \alpha (2 \go - 1) - 2(1 - \e)^{\go}(\go + \dg\ln(1 - \e)) - \nonumber\\ 
            &~&~~~~~~3(1 - \e)^{1 - \go}(\go+ \dg\ln(1 - \e) - 1)~,\label{f1}\\
    f_2(\e) &=& 2\alpha \dg - 3(1 - \e)^{1-\go} \times \nonumber\\
&~& \left( 
\dg + \frac{1+(- \go+\ddg \ln(1 - \e))-(-1+\go + \dg \ln(1 - \e))^2}{2} \right) \nonumber\\
&-& 2(1 - \e)^{\go}
\left( \dg + \frac{- \go+\ddg \ln(1 - \e)+(\go + \dg \ln(1 - \e))^2}{2} \right)~.\label{f2}
\eea
We first solve for $\dg$, using (\ref{f1}) and we obtain 
\bea
\dg=\frac{(1 - \e)^{\go} (2 \go (1 - \e)^{\go}+ 3(\go - 1)(1 - \e)^{1 - \go} - \alpha (2\go - 1))}
                           {\ln(1 - \e)(3\e - 2(1 - \e)^{2 \go} + 2 \alpha (1 - \e)^{\go} - 3)}~.
\eea
Expanding this expression in series of $\e$ using \eqref{serieg} then gives
\begin{eqnarray}
\dg=\frac{\alpha}{5(2\alpha-5) \e}- \frac{26\alpha^2-5\alpha+150}{250(2\alpha-5)^2} 
                       +  \hdots \lb{dg2}
\end{eqnarray}
We finally use \eqref{dg2} in $f_2$, \eqref{f2}, in order to solve for $\ddg$ and we obtain the 
following expansion (the closed form expression is too complicated to be of interest)
\begin{align}
    \ddg &=\frac{2\alpha}{5(2\alpha - 5)\e^2}
-\frac{\alpha(504 \alpha^2 - 1525\alpha + 1250)}{125(2\alpha - 5)^2 (4\alpha - 5) \e} 
                                                                                  \nonumber\\
&+\frac{61696 \alpha^5 - 141960 \alpha^4 + 658150 \alpha ^3 - 2382375 \alpha^2 + 
                   3413125 \alpha - 1818750}{18750(2\alpha - 5)^3 (4\alpha - 5)^2}~.\lb{ddg2}
\end{align}
This is the main result of this Appendix.

Let us remark that near $\ot=1$, the derivative is given, up to terms of 
order ${\cal O}(\e){\cal O}(\go)$ by
\be
\frac{d\gamma}{d\ot}\simeq
\frac{(5 \go -3 )\e}{2 \alpha (1-\ot) \ln((1-\e)\ot)}
-\frac{\alpha (2 \go - 1) + 3- 5 \go}{2 \alpha \ln((1-\e)\ot)}~.
\ee
For $\e>0$, this expression is not singular at $\ot=1$ provided that the first numerator 
vanishes, i.e. $\go=\frac{3}{5}+{\cal O}(\e)$, as we have found in (\ref{pse}). 
For $\e=0$, the first term is identically zero and the condition for the derivative to be 
non singular at $\ot=1$ is $\g_{-\infty}=(\alpha-3)/(2\alpha-5)$ as obtained in (\ref{ginf}).

\end{document}